\documentclass[aps,prb,twocolumn,10pt,superscriptaddress,amsfonts,amssymb,amsmath,floatfix,floats,a4paper]{revtex4-1}

\usepackage{latexsym,epsfig,epstopdf}
\usepackage{graphicx}

\newcommand{\ket}[1]{\left | #1 \right\rangle}

\newcommand{\Tr}{\mathrm{Tr}}

\begin{document}

\title{Non-classicality of spin structures in condensed matter: An analysis of Sr$_{14}$Cu$_{24}$O$_{41}$}

\author{Wen Yu Kon}
\affiliation{School of Physical and Mathematical Sciences, Nanyang Technological University, 637371 Singapore, Singapore}

\author{Tanjung Krisnanda}
\affiliation{School of Physical and Mathematical Sciences, Nanyang Technological University, 637371 Singapore, Singapore}

\author{Pinaki Sengupta}
\affiliation{School of Physical and Mathematical Sciences, Nanyang Technological University, 637371 Singapore, Singapore}
\affiliation{MajuLab, International Joint Research Unit UMI 3654, CNRS, Universit\'e C\^ote d'Azur, Sorbonne Universit\'e, National University of Singapore, Nanyang Technological University, Singapore}

\author{Tomasz Paterek}
\affiliation{School of Physical and Mathematical Sciences, Nanyang Technological University, 637371 Singapore, Singapore}
\affiliation{MajuLab, International Joint Research Unit UMI 3654, CNRS, Universit\'e C\^ote d'Azur, Sorbonne Universit\'e, National University of Singapore, Nanyang Technological University, Singapore}
\affiliation{Institute of Theoretical Physics and Astrophysics, Faculty of Mathematics, Physics and Informatics, University of Gda\'nsk, 80-308 Gda\'nsk, Poland}

\begin{abstract}
When two quantum systems are coupled via a mediator, their dynamics has traces of non-classical properties of the mediator.
We show how this observation can be effectively utilised to study the quantum nature of materials without well-established structure.
A concrete example considered is Sr$_{14}$Cu$_{24}$O$_{41}$.
Measurements of low temperature magnetic and thermal properties of this compound were explained with long-range coupling of unpaired spins through dimerised spin chains.
We first show that the required coupling is not provided by the spin chain alone and give alternative compact two-dimensional spin arrangements compatible with the experimental results.
Then we argue that any mediator between the unpaired spins must share with them quantum correlations in the form of quantum discord and in many cases quantum entanglement.
In conclusion, present data witnesses quantum mediators between unpaired spins in Sr$_{14}$Cu$_{24}$O$_{41}$.
\end{abstract}

\maketitle

\section{Introduction}

Condensed matter physics concerns itself with materials that contain a large number of constituents with often unknown detailed atomic and spin structure.
Is it then possible to reveal quantum character of unknown arrangement of spins without elaborate modelling?
Here we provide affirmative answer by employing the method inspired by quantum communication scenarios.~\cite{revealing}
A conceptually simple application of this method would be to place the whole material in the role of a mediator of quantum entanglement between otherwise non-interacting quantum probes.
Intuitive reasoning relies on the fact that quantum entanglement cannot be created via local operations and classical communication.~\cite{ent-review}
Since the probes are coupled through the solid, the interactions are local and any entanglement gain witnesses non-classical communication.
The relevant non-classicality is captured by quantum discord that the solid must share with the probes.~\cite{hv2001,oz2002,discord-review}
In turn, such quantum discord is related to possibility of preparing non-orthogonal states of the material by operating on the probes only.~\cite{Streltsov2017,dis-pr}
In this way, from the observation of entanglement gain between the probes, one confirms quantum features of the material, without ever operating on it or making assumptions about its physics.

Here we show how to apply this idea differently.
Instead of placing the whole solid in the role of a mediator we identify both the probes and mediators within the sample. 
Our material of choice is the magnetic compound Sr$_{14}$Cu$_{24}$O$_{41}$.
We first argue that the spin structure in this solid is not yet fully understood by providing a number of spin arrangements compatible with experimental data.~\cite{Sahling2015}
They are all based on the existence of unpaired spins coupled via two-dimensional configurations of dimerised spin chains.~\cite{Gelle2004,GelleHAL}
Hence, we treat the unpaired spins as the probes and the dimerised spins of the two-dimensional spin arrangements as the mediators.
There are also other possible forms of mediators, for example the unpaired spins could be coupled via Ruderman-Kittel-Kasuya-Yosida mechanism.~\cite{rk1954,kasuya1956,yosida1957}
It is the advantage of our method that the details about the mediator and the probes are not important.
If a gain of quantum entanglement is observed between the probes, the coupling ``channel'' shares quantum correlations with them.
By introducing probes and mediators within the material it now becomes possible to make conclusions about their quantumness from measurements on the bulk, e.g. magnetisation or heat capacity.
This is a step forward compared to existing methods determining the presence of entanglement using bulk measurements~\cite{Wiesnaik2005,Das2011}
because additionally to entanglement, our method detects other forms of quantum correlations (discord) and it reveals which subsystems in the solid share quantum correlations.

We apply the method to the data of Sahling~\emph{et al.} who argued that unpaired spins of Sr$_{14}$Cu$_{24}$O$_{41}$ are entangled at low temperatures based on the measured susceptibility.~\cite{Sahling2015,Wiesnaik2005}
Accordingly, our analysis shows that their samples contain quantum coupling channels.
For further illustration we show the results of simulations with the dimerised spins of the two-dimensional spin arrangements as mediators.
It turns out they in fact share quantum entanglement, a stronger form of quantum correlations than discord, with the unpaired spins.

The paper is organised following our main message that the method of Krisnanda \emph{et. al.}~\cite{revealing} detects quantum correlations in materials without well-established structure.
We introduce the basic methodology in Sec.~\ref{SEC_ACB}. 
In Sec.~\ref{SEC_MATERIAL} we argue that the spin structure of Sr$_{14}$Cu$_{24}$O$_{41}$ is not yet fully understood by providing a number of distinct spin arrangements compatible with experimental data.~\cite{Sahling2015}
Finally, Sec.~\ref{SEC_MED} shows that the material must contain quantum mediators independently of their detailed physical fabric.

\section{Methodology}
\label{SEC_ACB}

Let us begin with the concepts of entanglement and discord.
A bipartite state is separable (not entangled) if it can be represented as a mixture of uncorrelated states:
\begin{equation}
\rho_{XY}^{\mathrm{sep}} = \sum_j p_j \rho_{X|j} \otimes \rho_{Y|j},
\label{EQ_SEP}
\end{equation}
where e.g. $\rho_{X|j}$ represents the density matrix of system $X$ and $p_j$ forms a probability distribution.
One way to quantify entanglement is by the distance between a generic state $\rho_{XY}$ and its closest separable state.~\cite{ree}
If this distance is measured by the relative entropy,
\begin{equation}
S(\rho||\sigma)=\Tr(\rho\log\rho) - \Tr(\rho\log\sigma),
\end{equation}
where $\rho$ and $\sigma$ are the two states of interest, the resulting measure $E_{X:Y}$ is called relative entropy of entanglement.
The distance approach can be generalised to include on equal footing other forms of correlations.~\cite{Modi2010}
In particular, a bipartite state is called quantum-classical if it admits the following decomposition
\begin{equation}
\rho_{XY}^{\mathrm{qc}} = \sum_y p_y \rho_{X|y} \otimes | y \rangle \langle y |,
\label{EQ_QC}
\end{equation}
where the kets $| y \rangle$ are mutually orthogonal.
The idea of classicality here is that there exists a von Neumann measurement on particle $Y$ which does not disturb the whole bipartite state.
The relative entropy distance $D_{X|Y}$ from a generic state $\rho_{XY}$ to the closest quantum-classical state is known as the relative entropy of discord or one-way quantum deficit.~\cite{Modi2010,deficit}
Since the set of states defined by Eq.~(\ref{EQ_QC}) is a subset of separable states, it follows that $D_{X|Y} \ge E_{X:Y}$, i.e. discord can be present in unentangled states.

In practice, relative entropy of entanglement is often hard to compute.
In our simulations we will therefore resort to negativity $N_{X:Y}$ being a computable entanglement monotone.~\cite{neg} Negativity of a quantum state $\rho$ is defined as
\begin{equation}
N_{X:Y}=\sum_{\lambda<0} |\lambda| ,
\end{equation}
where $\lambda$'s are the eigenvalues of $\rho^{T_X}$, the partial transpose of $\rho$ relative to system $X$.

Finally, we briefly recall the method introduced in Krisnanda \emph{et. al.}~\cite{revealing} with focus on its assumptions and possible applications to condensed matter systems.
The considered scenario involves two probes, $A$ and $B$, coupled via a mediator $C$.
This means that $A$ and $B$ do not interact directly, i.e. the total Hamiltonian is of the form
\begin{equation}
H = H_{AC} + H_{BC}.
\end{equation}
This is naturally satisfied for probes which are separated in space by a distance much larger than their position uncertainty, which is often the case in solids.
The second important assumption concerns environments of the considered subsystems.
Each of them is allowed to couple only to its local environment so that entanglement between the probes is not mediated via their common surroundings.
This has to be scrutinised carefully in each concrete application of the method or the common environment needs to be included in the mediator $C$.
Under these assumptions (and only these, note that the Hamiltonians can remain unknown as well as the initial state and dimensions of the involved Hilbert spaces) the following implication holds:
\begin{equation}
D_{AB|C}(t) = 0 \quad \Rightarrow \quad E_{A:BC}(\tau) \le E_{A:BC}(0).
\label{EQ_IMPL}
\end{equation}
That is, if the state of the $ABC$ system is quantum-classical (classical on the mediator) at all times $t \in [0,\tau]$, quantum entanglement $E_{A:BC}$ cannot grow.
Furthermore, since entanglement cannot increase under tracing out of a local subsystem $E_{A:B}(\tau) \le E_{A:BC}(\tau)$.
The same implication holds under exchange $A \leftrightarrow B$.
One can now estimate the initial entanglement $E_{A:BC}(0)$, but note that this requires access to the mediator.
Alternatively, this entanglement can be upper-bounded by operating on the probes only.
An example of such a bound is given by the sum of initial entropies of the probes, $E_{A:BC}(0) \le S_A(0) + S_B(0)$.~\cite{revealing}
Therefore, if at time $\tau$ one observes $E_{A:B}(\tau) > S_A(0) + S_B(0)$ then, according to Eq.~\eqref{EQ_IMPL}, there must have been non-zero discord $D_{AB|C}$ during the process.
We stress that the mediator is completely excluded from this assessment.

In practice estimation of the initial entanglement can come from the physics of studied system.
For example, if the initial temperature of the solid is high relative to the effective coupling strengths, it is reasonable to suspect negligible entanglement.
If at a later time, when the solid is cooled, entanglement between the probes is observed we conclude the presence of non-zero quantum discord $D_{AB|C}$ during the cooling process.
We emphasise again the minimalistic assumptions about the mediator.
In the example below we will apply this method to the compound where the mediator is at present even unknown,
but there is compelling evidence that mediated interactions indeed exist.

\section{Material}
\label{SEC_MATERIAL}

\begin{figure}
	\includegraphics[width=0.45\textwidth]{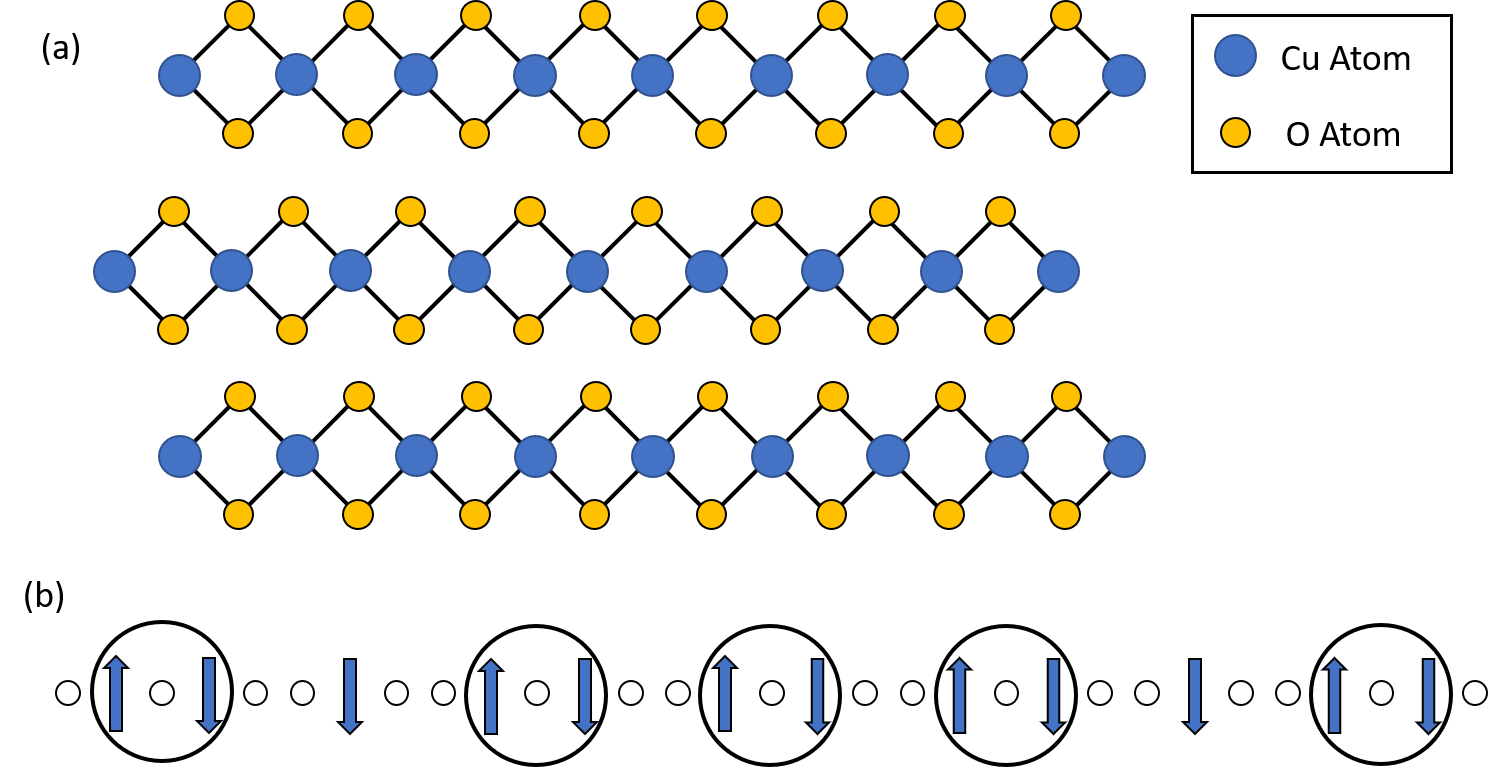}
	\centering
	\caption{Spin chains of Sr$_{14}$Cu$_{24}$O$_{41}$. 
	a: The spin chain sub-lattice is made up of CuO\textsubscript{2} chains.
	Each Cu atom contains an unpaired electron and hence is spin $\frac{1}{2}$.
	b: Due to the presence of holes in the chains most Cu atoms form Zhang-Rice singlets denoted by small empty circles.
	Experiments show that the structure is organised having spin dimers with exchange interaction $J_1 = 115$~K (intra-dimer coupling) for spins separated by one Zhang-Rice singlet.
	We denote these dimers with big black circles.
	The spins separated by two Zhang-Rice singlets are coupled via $J_2 = -13$~K (inter-dimer coupling).
	Spins on neighbouring chains are coupled with exchange interaction $J_{\perp} = 20$~K.~\cite{Regnault1999,Matsuda1999,Sahling2015}
	}
	\label{FIG_MAT}
\end{figure}

Sr$_{14}$Cu$_{24}$O$_{41}$ is composed of alternating two-dimensional layers of spin chains and spin ladders, with a layer of Strontium atoms in between, see, e.g. Fig.~1 in Blumberg \emph{et. al}.~\cite{blumberg2002}
The unit cell parameters of the alternating layers, in the direction of both chains and ladders, are incommensurate, with pseudoperiodicity of ten chain units for seven ladder units.
Here we focus on the spin chain sub-lattice whose structure is explained in Fig.~\ref{FIG_MAT}.
In short, each spin chain is composed of dimers with intra-dimer coupling $J_1 = 115$~K and inter-dimer coupling $J_2 = -13$~K, and a small number of unpaired spins.
\emph{Ab initio} calculations by Gell{\' e} and Lepetit show that the incommensurability of the chain sub-layer and the ladder sub-layer gives rise to the unpaired spins being separated by three to four spin dimers.~\cite{Gelle2004,GelleHAL}
Although questions remain about the actual number of holes in the chain sub-layer, many experiments point to the presence of spin dimers with gap around $115$~K and small amounts of unpaired spins.~\cite{Matsuda1996,Takigawa1998,Bag2017,Klingeler2006,Huvonen2007}

\begin{figure}
	\includegraphics[width=0.48\textwidth]{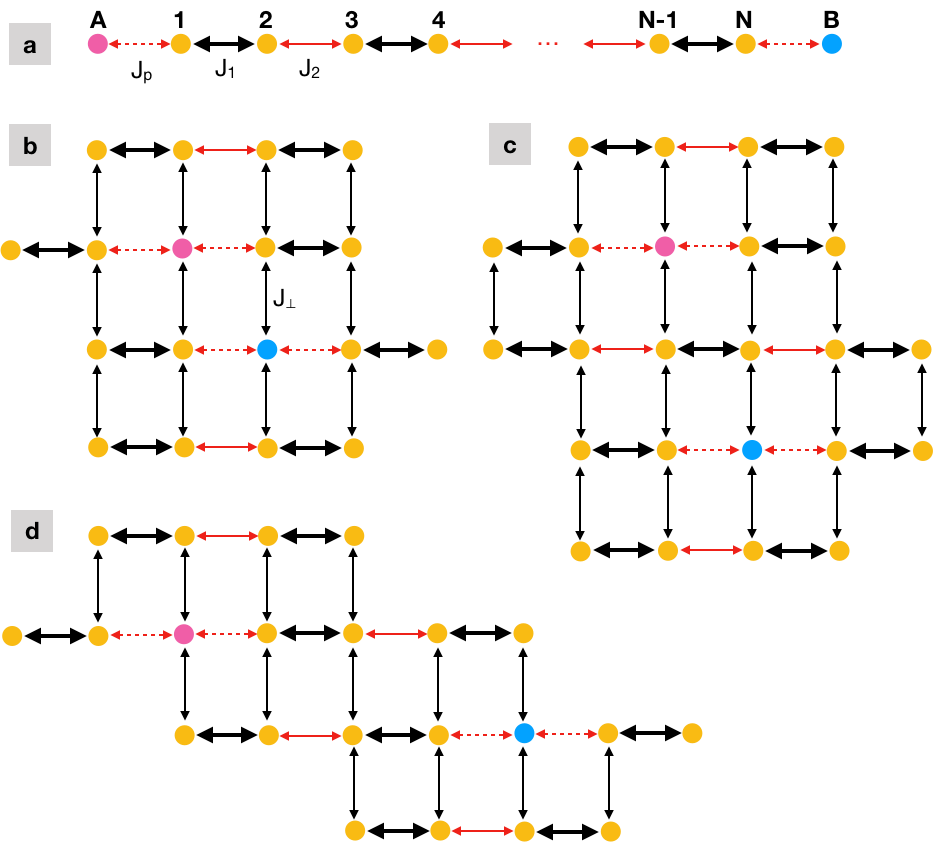}
	\centering
	\caption{Spin configurations in Sr$_{14}$Cu$_{24}$O$_{41}$. 
	a: One-dimensional spin chain model.
	Unpaired spins $A$ and $B$ are coupled through spin chain with alternating interactions $J_1 = 115$~K and $J_2 = -13$~K.
	The unpaired spins are coupled to the chain via interaction $J_p$ of unknown magnitude.
	We show that only the model with two spins in the chain matches the experimental results.~\cite{Sahling2015}
	b, c and d: Examples of two-dimensional configurations where nearby unpaired spins form effective dimers accounting for the experimental data ($J_\perp = 20$~K).
	The configuration with maximal distance between unpaired spins is presented in panel d.
	}
	\label{FIG_CHAINS}
\end{figure}

In a series of low temperature experiments, Sahling \emph{et. al.}~\cite{Sahling2015} measured magnetic and thermal properties of Sr$_{14}$Cu$_{24}$O$_{41}$ 
and noted that the observed data is compatible with the existence of new type of spin dimers having energy gap of about $2.3$~K.
We shall call them effective dimers.
The authors suggested that these low-temperature effective dimers could be formed {in the spin chain sub-lattice} from unpaired spins coupled by the dimerised spin chains as depicted in Fig.~\ref{FIG_CHAINS}a.
We now show that such a chain would have to be very short, in disagreement with the \emph{ab initio} calculations, and then provide alternative spin arrangements that match the $2.3$~K effective dimers observed in the experiment.

Consider the spin chain as in Fig.~\ref{FIG_CHAINS}a, where the coupling of the unpaired spins to the chain, $J_p$, is of unknown magnitude.~\cite{Sahling2015}
We first assume that $J_p \ll J_1$.
In this case we show that the effective coupling between the unpaired spins decays exponentially with the size of the spin chain
and there is no configuration matching the experimentally established energy gap of $2.3$~K.
Since $J_1$ is an order of magnitude higher than $J_2$ it is justified to utilise perturbation theory. 
We adopt the so-called decimation method, first proposed to simplify the problem of a random anti-ferromagnetic spin chain,~\cite{Dasgupta1980}
and as a result obtain an effective dimer between the unpaired spins $A$ and $B$ with coupling energy and energy gap given by (see Appendix~\ref{APP_DECIMATION}):
\begin{equation}
J_{AB} = \frac{J_p^2}{J_2} \left( \frac{J_2}{2 J_1} \right)^{N/2},
\label{EQ_JAB}
\end{equation}
where $N$ is the total number of spins in the chain.
This approximation works well and, e.g., for $J_p = J_2$, matches quite closely the results of exact diagonalisation and density matrix renormalisation group simulation of the total Hamiltonian on finite chains.
We have done the corresponding numerics up to $N = 38$ (the predicted spin chain length in Sahling \emph{et. al.}~\cite{Sahling2015}) and present the results in Table~\ref{TAB_DMRG}.
Note that with these parameters even the shortest chain composed of one dimer does not match experimental data.

\begin{table}[!b]
	\centering
	\begin{tabular}{c||c|c|c}
		\, \, $N$ \, \, & \, \, \, Exact \, \, \, & \, \, \, Decimation \, \, \, & \, \, \, DMRG \, \, \, \\
		\hline \hline
		$2$ & $0.616465$ & $0.734783$ & $0.616465$\\
		\hline
		$4$ & $-0.033539$ & $-0.041531$ & $-0.033539$\\
		\hline
		$6$ & $0.001811$ & $0.002347$ & $0.001811$\\
		\hline
		$8$ & $-0.000098$ &$-0.000133$ & $-0.000098$\\
		\hline
		$10$ & $0.000005$ & $0.000007$ &$0.000005$\\
		\hline
		$38$ & - & $2.55\times10^{-23}$ & $<10^{-10}$ \\

	\end{tabular}
	\caption{Comparison of effective coupling between unpaired spins separated by a dimerised chain of $N$ spins.
	The left column gives the results of exact diagonalisation.
	This can only be done for short chains.
	For longer chains we used the density matrix renormalisation group simulation the results of which are given in the right column.
	Analytical treatment using decimation method gives results presented in the middle column. (Units: K)}
	\label{TAB_DMRG}
\end{table}

It is worth noting that despite small gaps the ground state of the one-dimensional system gives rise to significant entanglement between the unpaired spins.
In fact the weak coupling $J_p \ll J_1$ results in a strong long-distance entanglement,~\cite{Bazhanov2018} but only at absolute zero temperature.
Similar long-distance entanglement at low temperatures were observed for other spin chains.~\cite{CV2006,CV2007,Bose2010}
The small effective dimer gap makes the long-distance entanglement susceptible to thermal fluctuations
and would be responsible for its disappearance at the experimentally realised temperatures,~\cite{Kuwahara2010} for most separations between unpaired spins.

Since it is hard to estimate the value of $J_p$, we vary its magnitude and sign keeping in mind that it cannot be much larger than $J_1$.
For if $J_p \gg J_1$, the unpaired spins would form independent dimers with their nearest neighbours in the chain, which cannot explain a $2.3$~K effective dimer.
This requirement combined with Eq.~\eqref{EQ_JAB} shows that the chain cannot contain more than $N = 4$ spins.
However, for $N=4$ the ground state of the effective dimer is a triplet and therefore such a system would get magnetised in small external magnetic fields, again in contradiction to the experimental findings (see Fig. 3b in Sahling \emph{et al.}~\cite{Sahling2015}).
We are therefore left with the sole possibility of unpaired spins coupled by the chain of $N=2$ spins.
Indeed the experimentally established gap of $2.3$~K is explained by taking $J_p = - 27.8$~K or $J_p = 20.4$~K, obtained with exact diagonalisation.
We point out that such spins are separated in space by less than $20$~\AA.
Note also that if the structure in Fig.~\ref{FIG_MAT}b is taken seriously we expect $J_p\approx J_2$ as both couplings are across two Zhang-Rice singlets.
Since none of the 1D configurations examined above has $J_p$ close to $-13$~K, they cannot be responsible for $2.3$~K effective dimers.

In search for alternative spin arrangements we consider the entire spin-chain layer instead of a single chain as the magnitude of the interchain coupling, $J_\perp$, is higher than the coupling $J_2$ within the one-dimensional spin chains.
Using exact diagonalisation, we calculate the energy gaps of various 2D configurations. 
Some concrete examples of unpaired spins leading to effective dimers with the energy gap matching the observed value are presented in Figs.~\ref{FIG_CHAINS}b-d.
The unpaired spins are surrounded by spin dimers as one expects in the bulk.
We find numerically that the effective dimer forms independently of the configuration of the surrounding spins.
We verified all 2D possibilities of placing unpaired spins on the spin-chain sub-lattice, compatible with $J_p$ being not much larger than $J_1$.
The configuration with the most distant unpaired spins is presented in Fig.~\ref{FIG_CHAINS}d, where they are separated by $30$~\AA.
This requires $J_p = 75.1$~K, and is hence unlikely if one expects $J_p$ to be close to $J_2$.
Smaller couplings to the chain are possible at the expense of less diluted dimers.
The configuration presented in Fig.~\ref{FIG_CHAINS}b matches the data for $J_p = -14.3$~K, which obeys $J_p\approx J_2$. 
Variant~\ref{FIG_CHAINS}c requires $J_p = 11.5$~K.
However, the gap of this effective dimer does not change significantly with $J_p$.
In particular, the gap is $2$~K when $J_p=-13$~K.

We note that such 2D arrangements could occur in Sr$_{14}$Cu$_{24}$O$_{41}$ within the model similar to Fig.~\ref{FIG_MAT}.
According to this model each 1D chain contains unpaired spins separated by spin dimers.
These unpaired spins are expected to be homogeneously distributed in the chain since there is strong Coulomb force present.~\cite{Schnack2005} 
Sahling \emph{et. al.},~\cite{Sahling2015} found the concentration of one unpaired spin per $100$ Cu atoms, which corresponds to about $20$ dimer separation between the unpaired spins.
Since the atoms on neighbouring chains are shifted, see Fig.~\ref{FIG_MAT}a, it is natural to expect that the unpaired spins are also shifted, giving rise to the regular arrangements of the structure similar to that in Fig.~\ref{FIG_CHAINS}b.
The two unpaired spins on neighbouring chains would form the effective $2.3$~K dimers and the effective dimers would have little interaction with each other since they are spaced far apart.

The 2D spin configurations discussed point to the possibility that the effective dimer range may not be as large as previously thought.
However, how exactly the effective dimers are formed still remains an open question.
Furthermore, as we already hinted in the introduction, there might be yet another mechanism in operation producing effective dimers at low temperature.
Recent results,~\cite{Bag2018} introducing impurities into the material, support low-temperature effective dimers that are well separated.
A theoretical insight into further plausibility of our candidate spin configurations could be obtained from \emph{ab initio} calculations similar to those by Gell{\' e} and Lepetit,~\cite{Gelle2004}
but with the aim of determining the structure of the whole 2D chain sub-layer.
This is beyond the scope of the present work.
Moreover, there are suggestions that the unpaired spins observed in Sr$_{14}$Cu$_{24}$O$_{41}$ are actually trimers.~\cite{Klingeler2006} 
Our method remains applicable also in this case as we can take the probes A and B as the trimers.
Independently of the fabric of the mediator and probes, we now demonstrate that it has to share quantum correlations with the probes.

\section{Quantum mediators}
\label{SEC_MED}

\begin{figure*}
	\includegraphics[width=\textwidth]{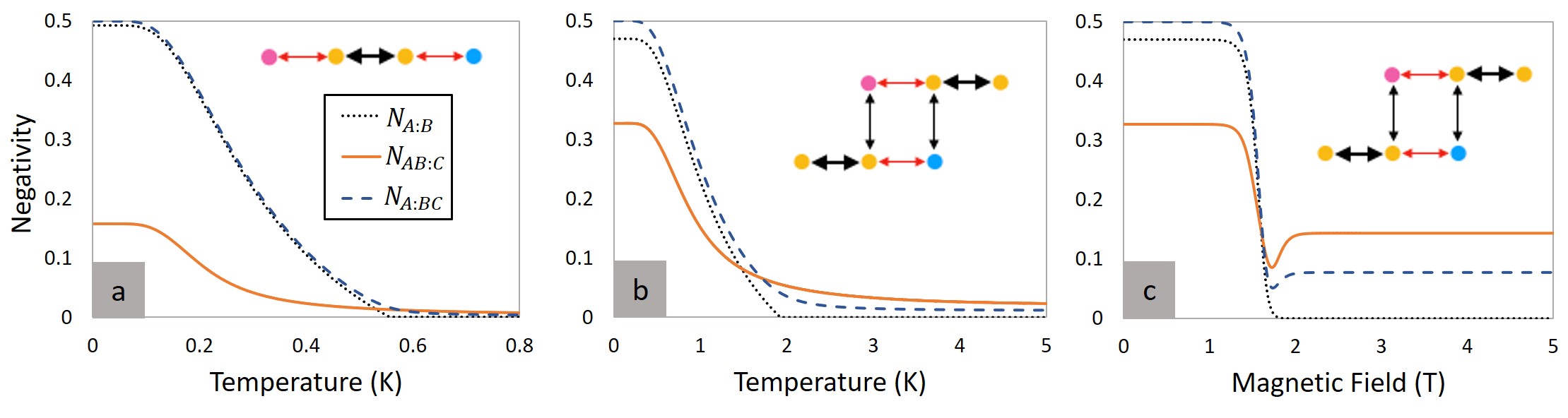}
	\centering
	\caption{Entanglement with mediators revealed by entanglement between unpaired spins. 
	The spin configurations for which entanglement is computed are given in the insets.
	Meaning of the symbols is the same as in Fig.~\ref{FIG_CHAINS}.
	In panels a and b negativity is computed as a function of surrounding temperature.
	For temperatures above $4$~K the unpaired spins are practically disentangled from the rest of the system (blue dashed line depicting $N_{A:BC}$).
	As we lower the temperature the unpaired spins form an effective dimer with high amount of entanglement (black dotted line showing $N_{A:B}$).
	This entanglement is always accompanied by entanglement between the unpaired spins and the mediator (yellow solid line giving $N_{AB:C}$).	
	In panel c we present the negativity as a function of external magnetic field (temperature is set to $0.1$~K).
	Also in this scenario entanglement between probe unpaired spins $N_{A:B}$ always witnesses entanglement with the mediator $N_{AB:C}$.
	}
	\label{FIG_SIM}
\end{figure*}

In order to apply the methodology for witnessing non-classicality we identify two unpaired spins with the probes $A$ and $B$ and the objects that couple them with the mediator $C$.
For example, in Fig.~\ref{FIG_CHAINS}a the mediator is the spin chain and in Figs.~\ref{FIG_CHAINS}b-d it is the respective two-dimensional arrangement of spins.
Indeed in all these cases the unpaired spins are not directly coupled.
Magnetisation and heat capacity data of Sahling~\emph{et al.} shows formation of the effective dimers at low temperature.
We therefore take as initial condition the experimental setting with high temperature.
As will be clear from the simulations below this is even satisfied at the temperature of liquid Helium.
In such a case the initial entanglement between any unpaired spin and the rest of the solid is vanishingly small.
By lowering the temperature to a value much smaller than the energy gap the effective dimers form between the unpaired spins. 
Their quantum state is close to the highly entangled ground state.
We conclude that quantum discord with the mediator must be created in the process. 

This conclusion is robust in a sense that it is valid when entanglement increases, independently of the couplings of unpaired spins to their mediator and local environments.
However, if more information is available stronger conclusions can be drawn.
We now demonstrate using the basic spin structures presented in Figs.~\ref{FIG_CHAINS}a-b that significant quantum entanglement 
(stronger form of quantum correlation than discord) with the mediator is also developed during cooling.
To this aim we present in Figs.~\ref{FIG_SIM}a-b negativity of thermal states corresponding to the spin configurations depicted in the insets.
In particular, finite entanglement $N_{A:B}$ between the unpaired spins is always accompanied by entanglement with the mediator $N_{AB:C}$.

These conclusions are reached without direct measurements of entanglement.
Instead they are a consequence of a combination of the results of bulk measurements on the whole solid.
The unpaired spins have been observed in other experiments,~\cite{Takigawa1998} and magnetisation and heat capacity data is compatible with small entanglement at high temperature (free spin behaviour) 
and entanglement gain between unpaired spins at low temperature (formation of effective dimers).~\cite{Sahling2015}

Another experimentally-feasible way to determine quantum features of the mediator would be to vary external magnetic field at constant temperature.
The presence of external magnetic field removes energy degeneracy of excited triplet subspace of the effective dimer and leads to a product ground state at sufficiently high external fields, i.e., the unpaired spins align with strong external field.
As the magnitude of the field gets diminished entanglement between the unpaired spins forms, thus witnessing quantum discord with the meditator.
This effect can be observed from the low-temperature experiments performed by Sahling \emph{et. al.}, where magnetisation is high at high fields, and decreases significantly at a critical magnetic field.~\cite{Sahling2015}
This is indicative of triplet-singlet transition of a dimer and thus one can conclude entanglement gain of the unpaired spins.
Again, simulations of this process with concrete parameters of Sr$_{14}$Cu$_{24}$O$_{41}$ show that entanglement in the effective dimer is always accompanied by entanglement with the mediator, see Fig.~\ref{FIG_SIM}c.

Simulation in external magnetic field has interesting features that we wish to discuss.
First of all, a finite entanglement with mediator $N_{AB:C}$ and $N_{A:BC}$ is present even at higher field strengths.
At low temperatures ($T = 0.1$~K), thermal fluctuations above the ground state are negligible.
Our results show that the ground state spin configuration remains the same for all fields in the range $2\lesssim B<90$ T, despite the coupling strength of the unpaired spins to the mediator being $20$ K (about $15$ T) in this simulation.
The spin system becomes disentangled only when the fields are strong enough to break the intra-dimer coupling $J_1 = 115$~K, i.e. above $90$ T.
The second observation is a dip in entanglement with mediator at magnetic field close to $1.7$~T.
At this field the system experiences a transition between highly entangled dimer singlet ground state and a new far less entangled ground state of the whole system.
During this transition, the two states are close in energy and their thermal mixture is the origin of the dip.

\section{Conclusions}

We have applied the method for revealing quantum correlations with inaccessible objects to condensed matter spin systems.
The main advantage of this method are minimalistic assumptions about the mediator itself.
The mediator does not even have to be well defined.
We argued that this is the situation in our present understanding of the structure of Sr$_{14}$Cu$_{24}$O$_{41}$.
Its low temperature magnetic and thermal properties are compatible with formation of an effective spin dimer.
By assuming that these dimers are formed by unpaired spins we showed a number of two-dimensional spin arrangements compatible with the experimental data.
Independently of which one of them is actually realised in the material or whether there exists yet another mechanism behind the formation of the effective dimers,
the discussed methodology shows that any mediator must share quantum discord with the unpaired spins.
Furthermore, simulations demonstrate considerable quantum entanglement between the unpaired spins and the two-dimensional spin mediators.
We hope this method will find many other applications in condensed matter.

\section{Acknowledgments}

We thank Emilio Lorenzo for useful discussions and experimental data.
This work is supported by Singapore Ministry of Education Academic Research Fund Tier 2 Project No. MOE2015-T2-2-034 and Tier 1 Project No. 2015-T1-001-056.
W.Y.K. thanks Nanyang Technological University for their support through the CN Yang Scholars Programme.

\appendix

\section{Decimation}
\label{APP_DECIMATION}

\begin{figure}[!b]
	\includegraphics[scale=0.5]{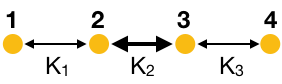}
	\centering
	\caption{The basic step in the decimation method replaces four spins $1-2-3-4$ with strong interaction in the middle, i.e., $K_2 \gg \max(K_1,K_3)$, with two spins, $1$ and $4$, coupled by effective interaction $K_1 K_3 / 2 K_2$.}
	\label{FIG_SI_CHAIN}
\end{figure}

Here we derive Eq.~(\ref{EQ_JAB}) of the main text.
For completeness, we first rigorously reproduce the result of Dasgupta and Ma.~\cite{Dasgupta1980}
Consider the situation depicted in Fig.~\ref{FIG_SI_CHAIN}.
The Hamiltonian of the four-spin system reads:
\begin{equation}
H = K_1 \, S_1 \cdot S_2 + K_2 \, S_2 \cdot S_3 + K_3 \, S_3 \cdot S_4,
\end{equation}
where e.g. $K_1$ is the coupling energy between spins $1$ and $2$, with respective vector spin operators $S_1$ and $S_2$.
We assume that $K_2$ is much larger than the remaining two couplings and apply perturbation theory with the base Hamiltonian and perturbation taken as
\begin{eqnarray}
H_0 & = & K_2 \, S_2 \cdot S_3, \\
H_p & = & K_1 \, S_1 \cdot S_2 + K_3 \, S_3 \cdot S_4.
\end{eqnarray}
By solving the base Hamiltonian for an antiferromagnetic coupling $K_2$, one obtains a singlet ground state (eigenenergy: $- \frac{3}{4} K_2$) and three-fold degenerate triplet excited states (eigenenergy: $\frac{1}{4} K_2$).
Since the base Hamiltonian does not act on spins $1$ and $4$ the ground state is degenerate with exemplary eigenstates:
\begin{eqnarray}
\ket{\phi_1} & = & \ket{\psi^-}_{23}\otimes\ket{\psi^-}_{14} \nonumber \\
\ket{\phi_2} & = & \ket{\psi^-}_{23}\otimes\ket{\psi^+}_{14} \nonumber \\
\ket{\phi_3} & = & \ket{\psi^-}_{23}\otimes\ket{\uparrow\uparrow}_{14} \nonumber \\
\ket{\phi_4} & = & \ket{\psi^-}_{23}\otimes\ket{\downarrow\downarrow}_{14}
\end{eqnarray}
where $\ket{\psi^\pm} = \frac{1}{\sqrt{2}} (\ket{\uparrow \downarrow} \pm \ket{\downarrow \uparrow})$.
In the perturbed system our focus is only on the low energy eigenstates as the gap to excited states, $K_2$, is much larger than the remaining couplings.
By performing second order degenerate perturbation theory we arrive at an effective Hamiltonian, coupling spins $1$ and $4$, given by
\begin{equation}
H_{14} = E + \frac{K_1 K_3}{2 K_2} S_1 \cdot S_4,
\label{EQ_H14}
\end{equation}
where $E = -\frac{3}{4} K_2 - \frac{3}{16 K_2} (K_1^2 + K_2^2)$ is the rigid energy shift that can be ignored.

The net result of decimation is that blocks of four spins with the only significant interaction in the middle can be replaced, as far as low energy is concerned, by two spins with effective coupling given in Eq.~(\ref{EQ_H14}).
We apply this method sequentially to the spin chain described in Fig.~\ref{FIG_CHAINS}a of the main text.
In the first step we replace spins $A-1-2-3$ by spins $A$ and $3$ only, with effective coupling $J_p J_2 / 2 J_1$.
Next, we replace spins $A-3-4-5$ by spins $A$ and $5$ only, with effective coupling $J_p J_2^2 / (2 J_1)^2$ etc.
Note that the number of spins in the spin chain, $N$, is always even.
By applying this procedure until we reach spin $B$ on the right hand side we arrive at Eq.~(\ref{EQ_JAB}).

\bibliographystyle{apsrev4-1-edit}
\bibliography{spins} 

\end{document}